\documentclass{optica-article}
\journal{oe}

\articletype{Research Article}

\usepackage{lineno}
\usepackage{siunitx}
\usepackage{mathtools}
\usepackage{array}
\usepackage{tabularx}
\usepackage{graphicx}
\usepackage{dcolumn}
\usepackage{bm}
\usepackage{amsmath}
\usepackage{color}
\usepackage{siunitx}
\usepackage{wasysym}
\usepackage{float}
\usepackage{array,url,kantlipsum}
\usepackage{verbatim}
\usepackage{textcomp}
\usepackage{caption}
\usepackage{xcolor}
\usepackage{subcaption}
\newcolumntype{P}[1]{>{\centering\arraybackslash}p{#1}}
\usepackage{array}
\usepackage{inputenc}
\usepackage[section]{placeins}
\usepackage{verbatim}
\DeclareSIUnit\gauss{Gs}

\begin{document}
\nolinenumbers
\title{Diamond-optic enhanced photon collection efficiency for sensing with nitrogen-vacancy centers }

\author{Muhib Omar\authormark{1,2,*}, Andreas Conta\authormark{2}, Andreas Westerhoff\authormark{2}, Raphael Hasse\authormark{2} , Georgios Chatzidrosos \authormark{1,2} 
, Dmitry Budker\authormark{1,2,3} and  Arne Wickenbrock\authormark{1,2}}

\address{\authormark{1}Helmholtz-Institut, GSI Helmholtzzentrum für Schwerionenforschung, Mainz, Germany\\
\authormark{2}Department of Physics, Mathematics and Computer Science, Johannes Gutenberg-Universität Mainz, Mainz, Germany\\
\authormark{3}Department of Physics, University of California, Berkeley, Berkeley, CA, United States}

\email{\authormark{*}momar@uni-mainz.de}

\begin{abstract}
We present a design to increase the amount of collected fluorescence emitted by nitrogen-vacancy color centers in diamond used for quantum-sensing.
An improvement was measured in collected fluorescence when comparing oppositely faced emitting surfaces by a factor of $3.8(1)$. This matches ray-tracing simulation results. This design therefore improves on the shot noise limited sensitivity in optical read-out based measurements of for instance magnetic and electric fields, pressure, temperature and rotations. \end{abstract}

%
%
%
%
\section{Introduction}
Nitrogen-vacancy centers (referred to as NV-centers in the following) are used in various applications ranging from high-precision temperature measurements \cite{2013toylithermometry}, magnetic-field measurements in various modalitites, with \cite{2015wolfsubpico,georgioscavity} or without employing microwaves \cite{arnemwfree} or bias fields\cite{tillzero} for instance, electric-field measurements\cite{doldeelectric} to quantum computing\cite{Quantumcomputer2010}, gyroscopy \cite{andreygyro} as well as bio-sensing \cite{barryworm}.

A NV-center is a point defect in diamond, where a pair of neighbouring carbon atoms is replaced by a nitrogen atom and a vacancy. It is an atom-like system, that can be in different charge states, called NV$^{+}$, NV$^{0}$ and NV$^{-}$, where the latter one is favoured for applications due to its optical spin read-out \cite{DOHERTY20131}. The NV$^{-}$ center has a total electron spin of 1, the corresponding electrons being contributed by the NV-center nitrogen itself, the open bonds from the carbon atoms and of another substitutional nitrogen atom in the lattice. The electron spins can be optically pumped into one of the NV's Zeeman sublevels by illuminating the diamond with, for example, \SI{532}{\nano \meter} laser light driving transitions into the phonon broadened excited state. The spin state can be read out by detecting the amount of (infra)red fluorescence light, due to spin selective non-radiative transitions \cite{DOHERTY20131}. Driving microwave transitions between the various spin states leads to observable changes in fluorescence. Those can be used to measure magnetic fields over the respective transition frequency shifts via the Zeeman effect . 

A fundamental noise limit of fluorescence detection arises from photon shot noise, depending on the amount of collected light which usually dominates over spin-projection noise as another fundamental limit. Therefore, by increasing the amount of collected light, the signal-to-noise ratio of such measurements can be improved.
Several different techniques have been developed to improve photon-collection efficiency, in both single-NV setups \cite{Haddena2010,Li2015,Choy2013} and ensemble experiments \cite{2015wolfsubpico,fourphotdiodes}.
\begin{figure}[h]
    \centering
    {\includegraphics[width=1\linewidth]{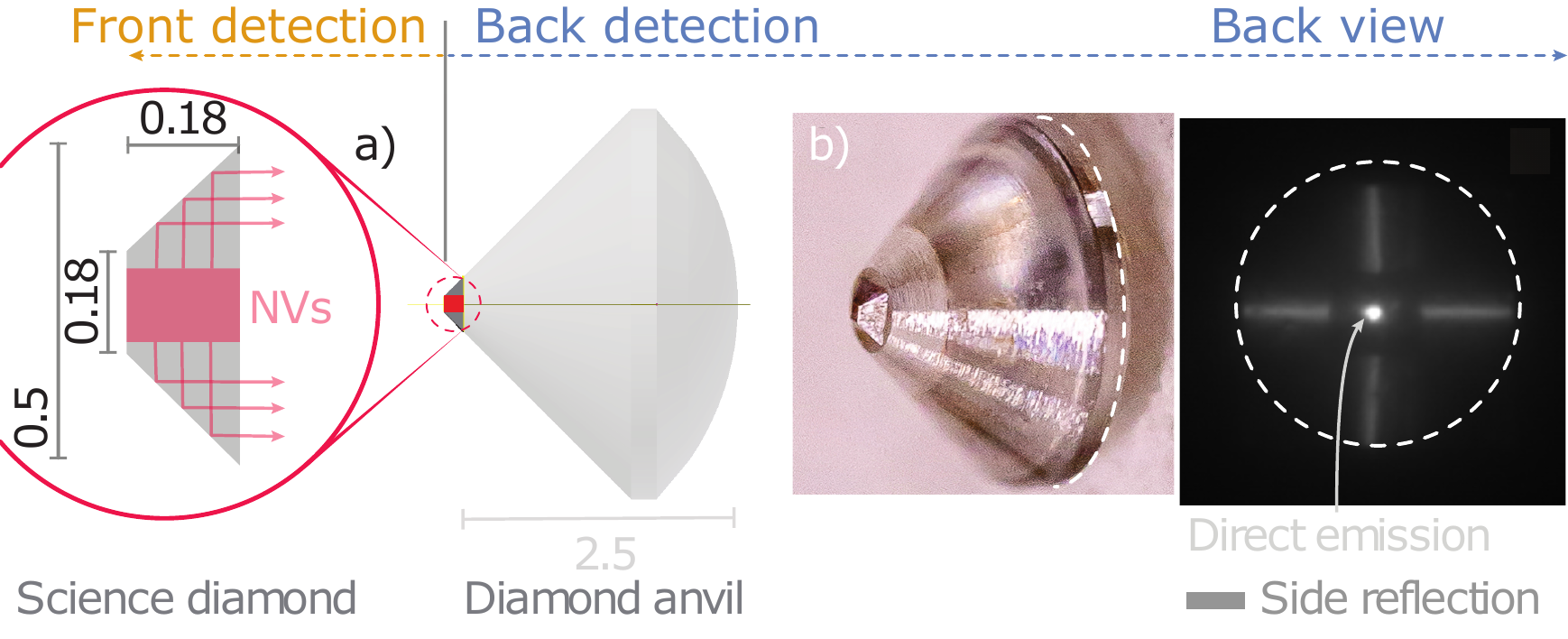}}
    \caption{Diamond assembly images: (a) Sketch of the NV-bearing diamond pyramid at the front (science diamond)  dimensions on top of the diamond anvil. (b) Photo of the science diamond, glued to the diamond anvil. c) Image of fluorescence light collected  from the back of the diamond anvil used for alignment purposes and taken with a CCD camera. The circle in the center of the cross is the apex of the fluorescence cone from the laser beam focal spot and the four side beams arise due to the side reflections of the anvil. All measures are in mm. }
    \label{fig:diamond}
\end{figure}

In this work, the diamond containing the NV centers, referred to as the sensing diamond, was glued to a cone-shaped diamond piece, referred to as the diamond anvil, which increases the amount of collected fluorescence light. The sensing diamond was cut to direct side emitted fluorescence from the sensing volume into the back direction via total internal reflection, see Fig.\,\ref{fig:diamond}. The curved back surface of the diamond anvil reduces losses due to total internal reflection at the diamond to air interface. Detection with a photodiode confirms an improvement factor of $3.8(1)$ expected from simulations compared to the other exiting surface.

\section{Sample preparation}
The sensing diamond, a  high-pressure high-temperature (HPHT) sample (Element Six DNV-B14), is specified to contain 13\,ppm nitrogen, 3.7\,ppm NV$^-$-centers and 1.4\,ppm NV$^0$-centers. This specific sample is $^{13}$C-depleted (99.999\% $ ^{12}$C). 
The sample was irradiated with \SI{5}{\mega\electronvolt} electrons at a dose of \SI{2e19}{\centi\meter^{-2}} and then annealed at \SI{700}{\degreeCelsius} for eight hours. Its measured minimal linewidth in a pulsed optically detected magnetic resonance (ODMR) experiment is around \SI{300}{\kilo\hertz}. 

 The shape of the diamond anvil and sensing diamond pieces was optimised using the COMSOL Multiphysics software.  The simulations were used to evaluate the improvement in fluorescence collection between the back and front side.
   
\begin{figure}[h]
    \centering
   \includegraphics[width=0.9\linewidth]{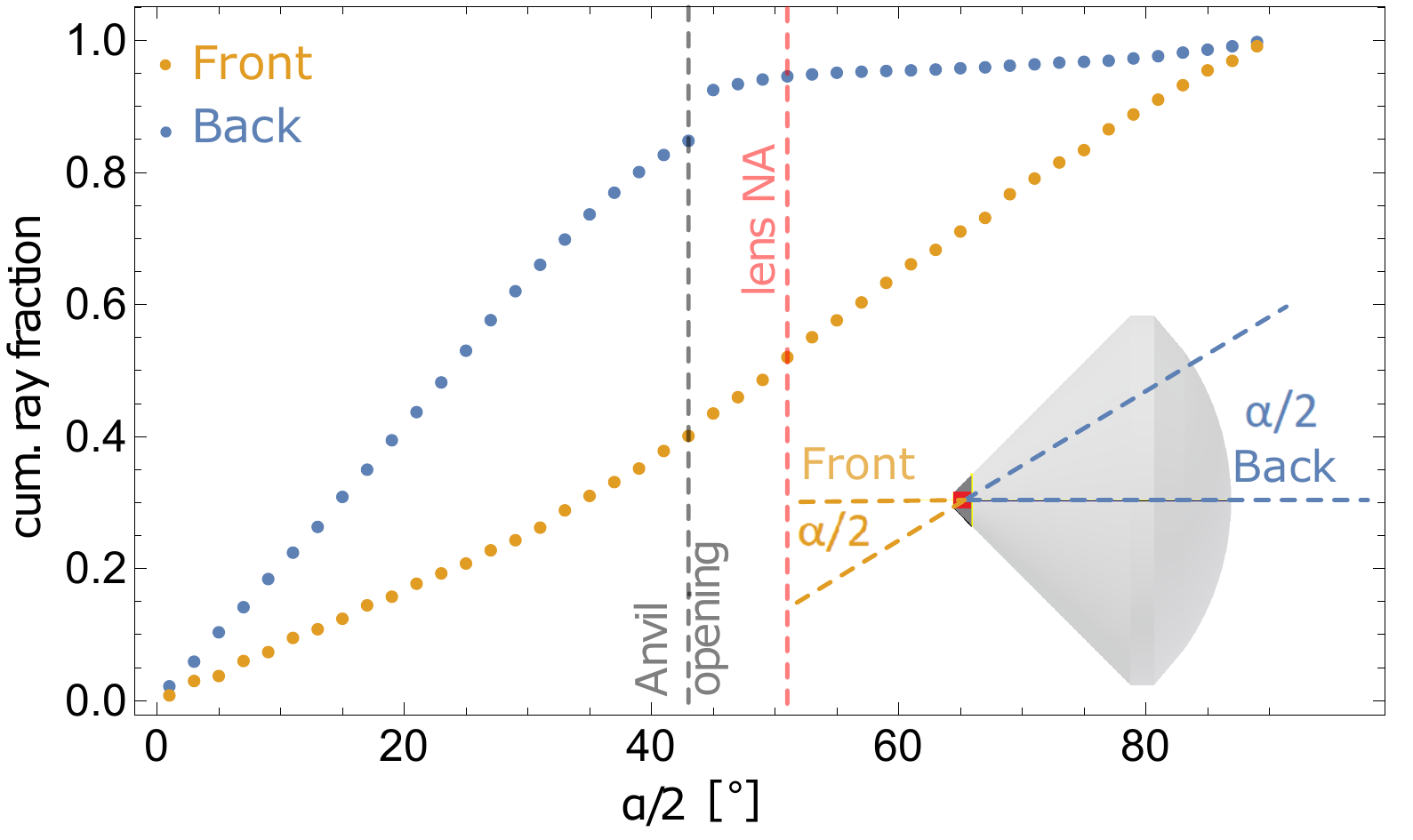}
    \caption{Simulated fractional cumulative angular ray distribution as a function of $\alpha$, the ray angle with respect to the symmetry axis of the diamond anvil with the sensing diamond. The cumulative ray fraction per single side collection is indicated as fraction of total number of emitted rays per side. The dashed red line indicates the numerical aperture of the collecting lens used in this note, the grey one the anvil opening angle of 45 degrees.  }
    \label{fig:Simulation}
\end{figure}

The science diamond is a trapezoid with a back surface being a square \SI{0.5}{\milli\meter} on the side, a height of \SI{0.18}{\milli\meter}, and the upper square surface being  \SI{0.15}{\milli\meter} on the side, see Fig.\,\ref{fig:diamond}. The base angle for this shape is close to 45 degrees to match with the single-crystal diamond anvil manufactured by Dutch Diamond Technologies. This limits the angular distribution of about 90\% of rays exiting the diamond construction to below 45 degrees with respect to the symmetry axis, see Fig.\,\ref{fig:Simulation}. This means 90\% of the light can be picked up by a lens with numerical aperture of 0.7. A very weak requirement. 
 The two diamond pieces were joined with a thin layer of Norland Optical Adhesive 170 with refractive index of 1.7  (the highest-index material that we could find) applied between the anvil and the back surface of the sensing diamond while pressing the pieces together. Effects as for instance etaloning due to a significant glue layer thickness were not observed.
 
 In the COMSOL simulation within the ray-tracing module, the number of collected rays were given a cylindrical distribution of ray sources inside the sensing diamond to mimic the excitation volume shape by the laser light inside the diamond. Three \SI{30}{\micro\meter} spaced point sources arranged along the symmetry axes of the sensing diamond emitted isotropically each 2000 rays. The ratio of collected rays between the front and back side on a photodiode using an 8 mm focal length, 12.7 mm diameter aspheric condenser lens was 3.8. 

 The diamond was mounted on a custom made peek mount during the experiment. Trying to test thermal durability of the glue joint, we applied around 1.8\,W of green laser light in a 0.9 mm diameter beam focused using 8\,mm focal length lens for around 10\,s on the sensing diamond from the back side. No degradation of the diamond optical assembly was observed at temperatures at which the peek material started to deform which are estimated to be around 150 °C based on its glass transition temperature \cite{peek}.

%
%
%
%
\section{Characterisation measurements}
To verify the simulation results we built a setup to measure the amount of fluorescence light collected from the front and back sides of the diamond simultaneously.  

\begin{figure}[h]
    \centering
    \includegraphics[width=0.9\linewidth]{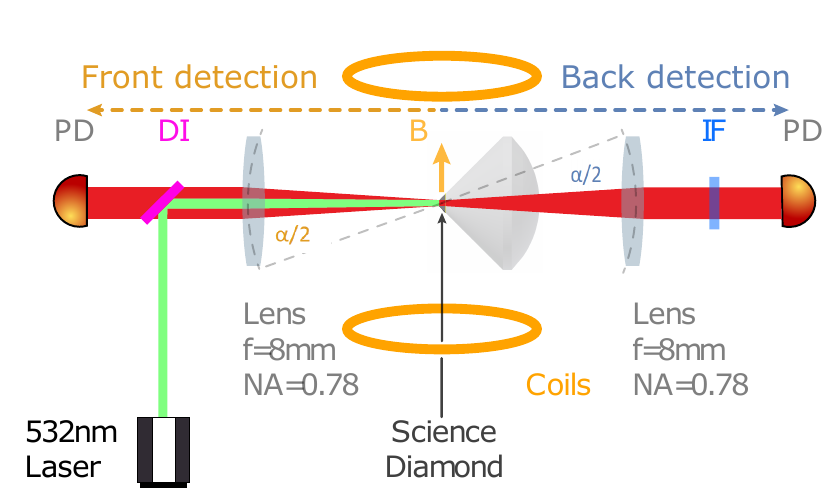}
    \caption[Experimental setup schematic]{Experimental setup: the \SI{532}{\nano \meter} laser light is focused on a NV-center doped sensing diamond, attached to a diamond anvil. The emitted light is focused onto the same model of photodiodes on each side. A dichroic mirror is used to block the reflected green light going towards the first photodiode, which is used to to record the intensity of the fluorescence light. An interference filter is employed to reject the laser light and transmit the fluorescence light. The right photodiode and lens were replaced with a CCD camera and a longer focal length lens respectively to capture images used to align the light beam with respect to the diamond.}
    \label{fig:experimental setup}
\end{figure}

\subsection{Experimental setup}
The setup is sketched in 
Fig.\,\ref{fig:experimental setup}. A \SI{532}{\nano\meter} laser beam was focused into the sensing diamond using a plano-convex lens with a focal length of $f=$ \SI{8}{\milli\meter}. Behind the diamond we placed initially another lens and a notch filter to separate out green light from the (infra)red fluorescence light emitted from the diamond detected with a charge-coupled device (CCD) camera. That way we were able to verify that the diamond was well centred illuminated with the \SI{532}{\nano\meter} light. The camera was positioned on the back side producing the characteristic cross shape shown in Fig.\,\ref{fig:diamond} (c). This shape originates from reflections of the side surfaces of the sensing diamond and allows for precise positioning of the diamond relative to the laser beam using a XYZ-stage.

After alignment, the optics at the back side were replaced with the same type of aspheric condenser lens with $f$=\SI{8}{\milli\meter} focal length, a notch filter and a photodiode. The fluorescence was compared in both front and back direction simultaneously. Integrated over the expected fluorescence spectrum the notch filter (Thorlabs NF 533-17) transmits about 2\%  more than the dichroic mirror  (Thorlabs DMLP-567 ), negligible within the measurement error.

\begin{figure}[h]
    \centering
    \includegraphics[width=0.7\linewidth]{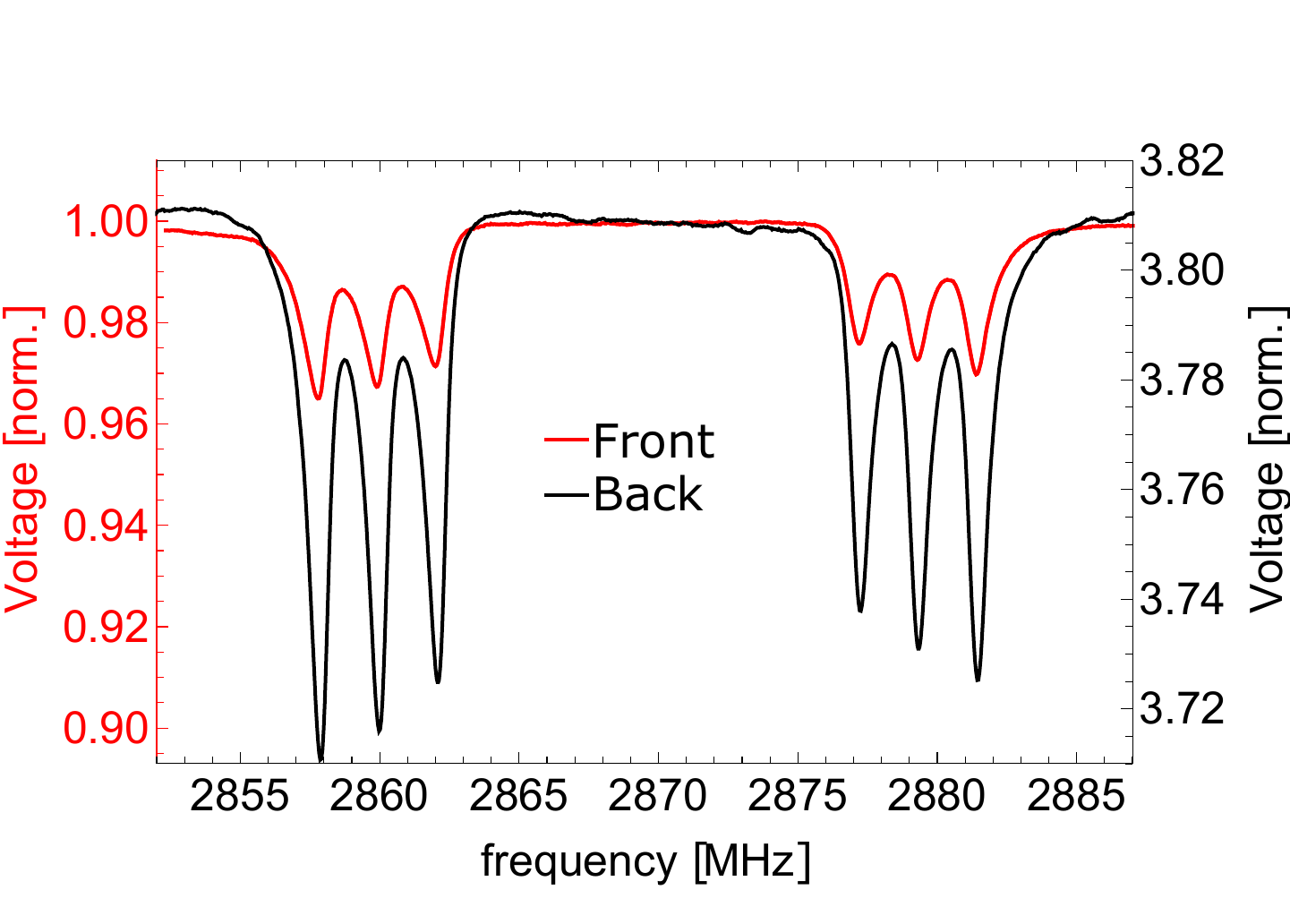}
    \caption{Comparison of the signal improvement due to the higher photon collection efficiency exemplified by an optically detected magnetic resonance spectra, both normalised to the back surface peak signal value.}
    \label{fig:odmr}
\end{figure}

\subsection{Measurements}

 Measuring on both sides simultaneously, for five laser powers equally spaced between 50 and 150\,mW gave a mean increase of collected fluorescence light by a factor of $ 3.8 (1)$ between the back and front side.
Next, a magnetic field was applied with Helmholtz coils and we used microwaves to obtain ODMR spectra of the NV centers to visualize the difference, see Fig.\,\ref{fig:odmr}.  

\section{Conclusion}
We described a design to improve the amount of collected fluorescence light emitted by a nitrogen-vacancy center ensemble in diamond. 
We were able to experimentally measure an increase by a factor of $3.8(1)$ between the improved design (back) with respect to the not improved opposing facet (front). This increase is supported by ray tracing simulations. An additional feature of the design is the improved angular distribution of the fluorescence. It would allow for over 90\% of the emitted fluorescence to be collected by a lens with a numerical aperture bigger than 0.7. These lenses are widely available. 
 In sensing applications relying on the collected fluorescence this improvement results in a lowered shot noise limit by a factor of nearly 2. Further improvement in overall light collection are possible for example deploying reflective coating on the front surface and anti-reflective coating on the back surface. Including the coating and neglecting losses this optic then allows to collect all emitted photons, which amounts to an additional increase of more than 40\%.  

\section{Funding}
This work was supported by the European Commission’s Horizon Europe Framework Program under the
Research and Innovation Action MUQUABIS GA no.10107054 and by the German Federal Ministry of Education and Research (BMBF) within the MILIQUANT project no. 13N15062 and DIAQNOS project no. 13N16455. 

\section{Acknowledgements}
We thank Dr. Till Lenz, Joseph Shaji Rebeirro and Omkar Dhungel for the many and fruitful discussions concerning this project.
\section{Disclosures}
The authors declare no conflicts of interest.
\section{Data availability}
Data underlying the results presented in this paper are not publicly available at this time but may be obtained from the authors upon reasonable request.

\end{document}